\documentclass{article}
\usepackage[final]{neurips_2023_ml4ps}

\usepackage[utf8]{inputenc} 
\usepackage[T1]{fontenc}    
\usepackage{hyperref}       
\usepackage{url}            
\usepackage{booktabs}       
\usepackage{amsfonts}       
\usepackage{nicefrac}       
\usepackage{microtype}      
\usepackage{xcolor}         
\usepackage{graphicx}
\usepackage{amsmath}
\newcommand\change[1]{{\color{black}#1}}

\title{Cosmological Field Emulation and Parameter Inference with Diffusion Models}

\author{%
  Nayantara Mudur \\
  Harvard University \\
  Cambridge, MA, 02138, USA \\
  \texttt{nmudur@g.harvard.edu}
  \And
  Carolina Cuesta-Lazaro \\
  The NSF AI Institute for Artificial Intelligence and Fundamental Interactions \\
  Massachusetts Institute of Technology, Cambridge, MA, 02139, USA \\
  \texttt{cuestalz@mit.edu} \\
  \And
   Douglas P. Finkbeiner \\
  Harvard University\\
  Cambridge, MA, 02138, USA \\
\texttt{dfinkbeiner@cfa.harvard.edu}
}

\begin{document}

\maketitle

\begin{abstract}
Cosmological simulations play a crucial role in elucidating the effect of physical parameters on the statistics of fields and on constraining parameters given information on density fields. We leverage diffusion generative models to address two tasks of importance to cosmology -- as an emulator for cold dark matter density fields conditional on input cosmological parameters $\Omega_m$ and $\sigma_8$, and as a parameter inference model that can return constraints on the cosmological parameters of an input field. We show that the model is able to generate fields with power spectra that are consistent with those of the simulated target distribution, and \change{capture the subtle effect of each parameter on modulations in the power spectrum}. We additionally explore their utility as parameter inference models and find that we can obtain tight constraints on cosmological parameters. 
\end{abstract}

\section{Introduction}
Cosmological simulations are expensive to run, and can only be generated for a limited set of initial conditions and points in parameter space. This has led to the need for emulators or surrogate models \citep{Heitmann:2009cu, mustafa2019cosmogan} that can learn to model the distribution of fields or summary statistics of importance to cosmology. A closely interlinked thrust of modern cosmology has been the search for statistics that can yield optimal constraints on cosmological parameters \citep{valogiannis2022going, dai2023multiscale}. Diffusion or score-based generative models \citep{song2020score} are generative models that involve a forward diffusion (noising) process. A neural network is then used to learn the denoising or generative process that maps samples from the standard normal to samples from the target distribution. The denoising diffusion probabilistic model (DDPM) \citep{ho2020denoising} formulation consists of a variance schedule ${\beta_t}$ over a fixed number of time steps, $T$, that controls the incremental noise added to the image, or the `speed' at which the target distribution is noised, and a score model, that is used to parameterize the reverse process. The model can also be conditional on either a discrete input label or a parameter vector. In this work, we apply diffusion generative models to emulate fields from cosmological simulations and \change{show that they can capture the dependence of summary statistics on input cosmology, and can provide tight constraints on cosmological parameters}.

\begin{figure}[h]
  \centering
\includegraphics[width=0.28\linewidth]{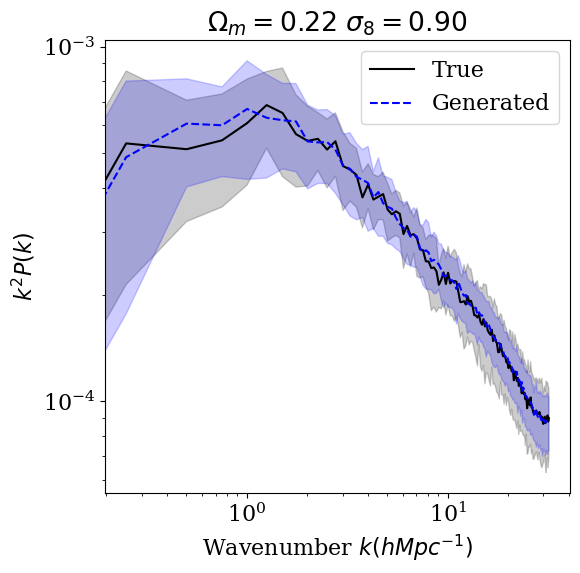}
\includegraphics[width=0.28\linewidth]{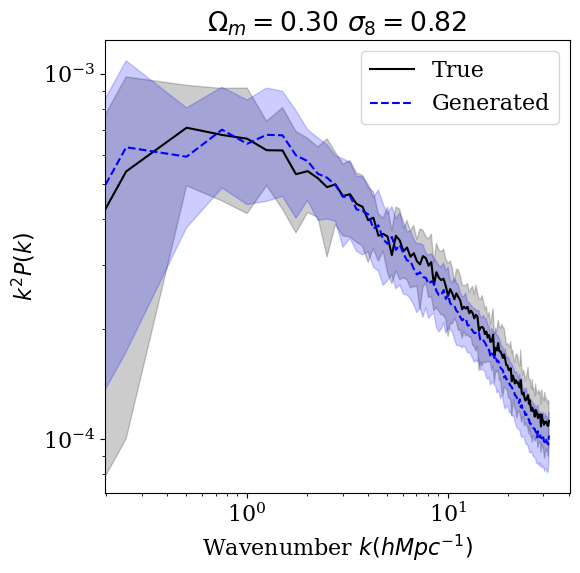}
\includegraphics[width=0.28\linewidth]{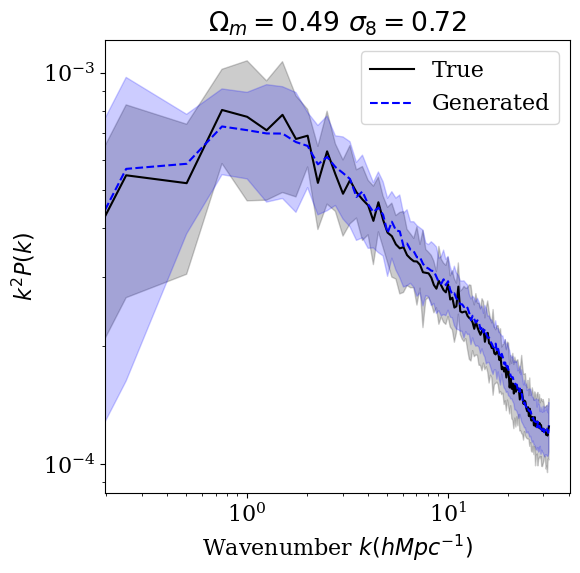}
\includegraphics[width=0.28\linewidth]{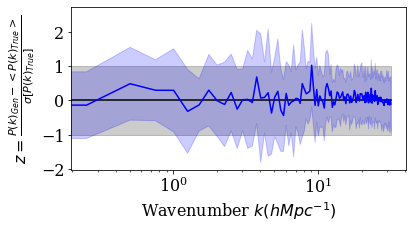}
\includegraphics[width=0.28\linewidth]{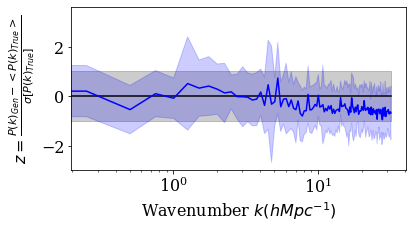}
\includegraphics[width=0.28\linewidth]{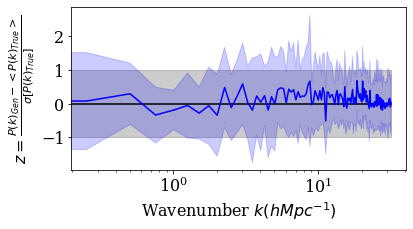}
\caption{$k^2P(k)$ (first row) and the \textit{z}-score (second row) for 3 different validation parameters.}
\label{fig:pk}
\end{figure}

\section{Datasets, Architecture and Training}\label{sec:training}
\paragraph{Datasets} We used the IllustrisTNG \citep{nelson2019illustristng, pillepich2018simulating} Cold Dark Matter density fields at $z=0$ from the CAMELS Multifield Dataset (CMD) \citep{villaescusa2022camels,villaescusa2021camels}. The dataset consists of 15 two-dimensional fields each for 1000 different points in parameter space. The parameter vector has 2 cosmological ($\Omega_m$ and $\sigma_8$) and 4 astrophysical parameters. The fields span $25 h^{-1}$ Mpc on each side. We work with the log (base 10) of these fields and randomly rotate or flip the field to account for invariance under rotations or parity.

\paragraph{Diffusion Model Setup} 
\newcommand\xt[1]{\mathbf{x}_{#1}}
\newcommand\qt[1]{q(\xt{#1}|\xt{#1 -1})}
\newcommand\pt[1]{p_\theta(\xt{#1-1}|\xt{#1})}
The forward diffusion process follows a variance schedule \{$\beta_t$\}, and $q(\xt{t}|\xt{0}) = \mathcal{N}(\sqrt{\bar{\alpha_t}}\xt{0}, (1 - \bar{\alpha_t})\mathbf{I})$, where $\bar{\alpha_t} = \prod_{t'=1}^t 1 - \beta_{t'}$.
The diffusion model architecture is similar to the U-Net \citep{ronneberger2015u} used in \cite{ho2020denoising}, has 4 down and up-sampling blocks consisting of 2 ResNet blocks \citep{zagoruyko2016wide}, group-normalization \citep{wu2018group}, and attention \citep{vaswani2017attention, shen2021efficient}. We use circular convolutions in the downsampling layers since the input fields have periodic boundary conditions. The model is conditional on $\Omega_m$ and $\sigma_8$. Each parameter is normalized to lie between [0, 1] with respect to its range, $\Omega_m \in [0.1, 0.5], \sigma_8\in [0.6, 1.0]$. The model has a multilayer perceptron that is used to transform the parameters into a space with the same dimension as the time embedding, and each ResNet block additionally has an MLP that takes cosmology as an input. We used a non-linear variance schedule with 1000 timesteps. During training, for each image $\xt{0}$, a timestep is sampled uniformly along with a noise pattern $\epsilon\sim\mathcal{N}(0, \mathbf{I})$. The loss function that is minimized is $L_{t-1} = ||\mathbf{\epsilon} - \mathbf{\epsilon}_\theta(\sqrt{\bar{\alpha}_t}\xt{0} + \sqrt{1 - \bar{\alpha}_t}\mathbf{\epsilon}, t, \vec{y}))||^2$, where $\vec{y}$ is the parameter vector. We used the Weights and Biases framework \citep{wandb} for our experiments. 
\paragraph{Training} We first train the conditional diffusion model on downsampled 64x64 images for 60000 iterations. To train the conditional diffusion model on 256x256 images, we initialize the architecture with the weights of the 64x64 model after 60000 iterations, and then train the model for over 500000 iterations. In our experiments, initializing the 256x256 model with the weights of the 64x64 model appeared to lead to faster convergence. The diffusion model loss is not informative of sample quality, and we need an alternative metric to assess the quality of the generated samples \cite{theis2015note}. We sample 50 fields for 10 different validation parameters for 4 different checkpoints -- corresponding to 200k, 220k, 240k and 260k iterations. For each checkpoint we compute the reduced chi-squared statistic of the power spectrum of each generated field, $s$, with respect to the power spectra of the 15 true fields from the dataset, corresponding to that parameter: $\chi_r^2 (s) = \frac{1}{|k|-1}\sum_k\frac{(P(k)_s - <P(k)_{\textsc{True}}>)^2}{\sigma[P(k)_\textsc{True}]^2}$. We then compute the mean of these values across all parameters and sampled fields. The checkpoint corresponding to the 260k iteration had the lowest value, corresponding to 1.30. \change{To put this number in perspective, we can examine the effect of cosmic variance on this metric using a leave-one-out cross-validation approach, by computing the reduced chi-squared statistic of each sample of a \textit{true} field, using the 14 other true fields corresponding to the same parameter as the reference distribution. The mean of this value across the 10 parameters is 1.27. This is consistent with the value of 1.27 that is obtained if the reduced chi-squared statistic is computed in a leave-one-out fashion using 15 observations of independently distributed Gaussians of length 128 (the number of bins in the power spectrum).} We use the 260k checkpoint for our analysis.

\section{Summary Statistics}
\begin{figure}[h]
  \centering
\includegraphics[width=\linewidth]{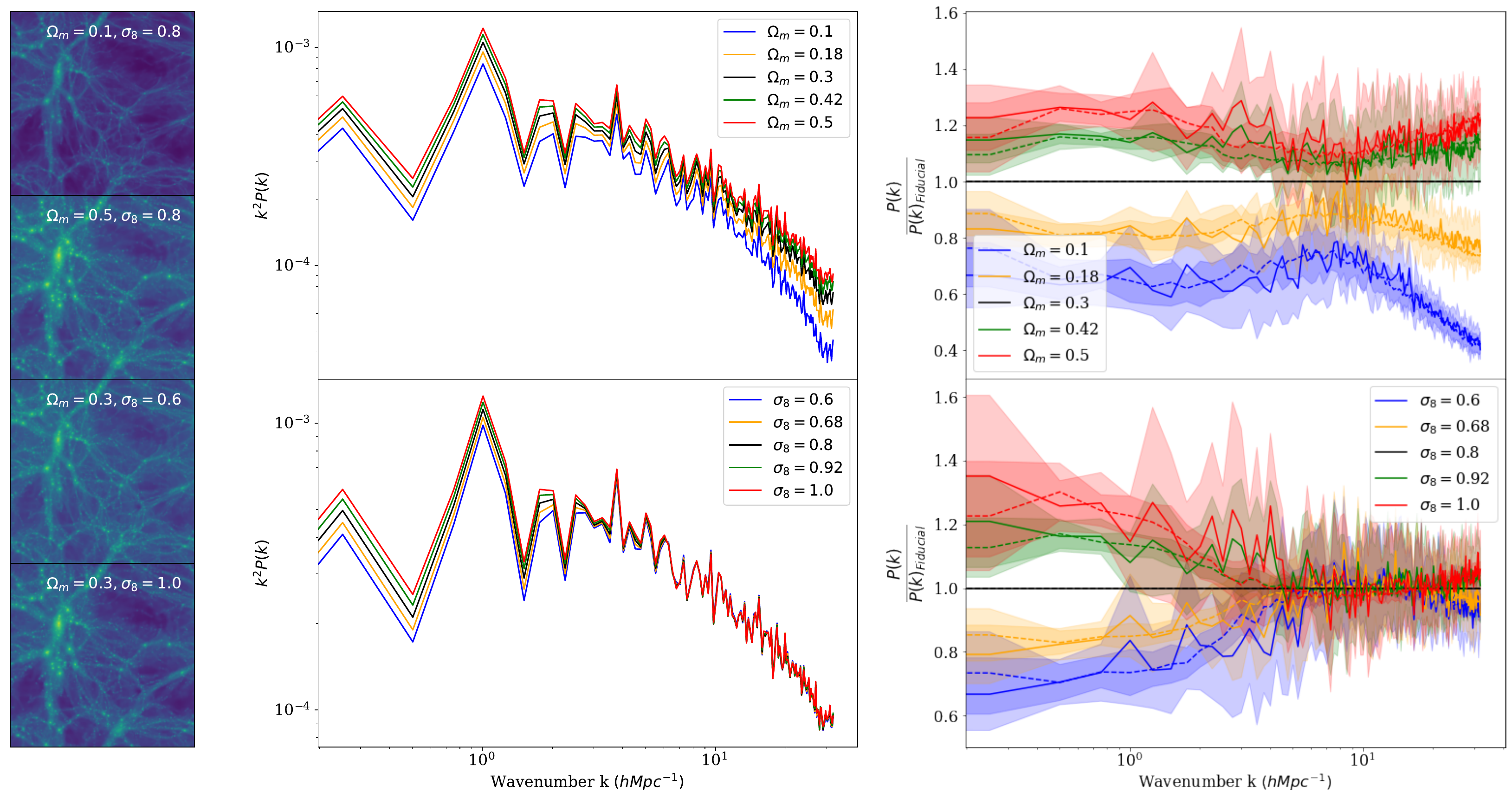}
\caption{Generated `1P' fields. \textbf{Left column:} Generated fields corresponding to the extreme values of each parameter for a single seed, with the other value held fixed at the fiducial value (0.3 for $\Omega_m$ and 0.8 for $\sigma_8$). \textbf{Middle column:} Power spectra of the generated fields for the same seed, for different values of each parameter, holding the other fixed. \textbf{Right column:} Mean and standard deviation for the ratio of the power spectra at the modified parameter value to the power spectra for the field at the fiducial parameter value (black) for 15 slices from the CAMELS dataset (solid) and 15 seeds for the generated fields from the diffusion model (dashed).}
\label{fig:conditional}
\end{figure}

We examine the power spectra for 3 different validation parameters for 15 true fields and 15 generated fields from the diffusion model, along with the $z$-scores in each $k$ bin in Figure \ref{fig:pk}. All statistics in this section are computed on the log of the true fields and the generated fields. \change{In Figure \ref{fig:conditional}, we generated `1P' sets and examined whether the effect of changing each parameter, while keeping the others constant is the same as is observed in the 1P CAMELS suite. Increasing $\Omega_m$ enhances the power spectrum at all scales in the generated fields (center) and affects the pixel values. Changes in $\sigma_8$ affect only scales larger than a few $h^{-1}$ Mpc. The one sigma envelope for the ratios of the modulations for the true fields appears to be slightly larger than that for the generated fields.}

\begin{figure}[!h]
  \centering
\includegraphics[width=0.35\linewidth]{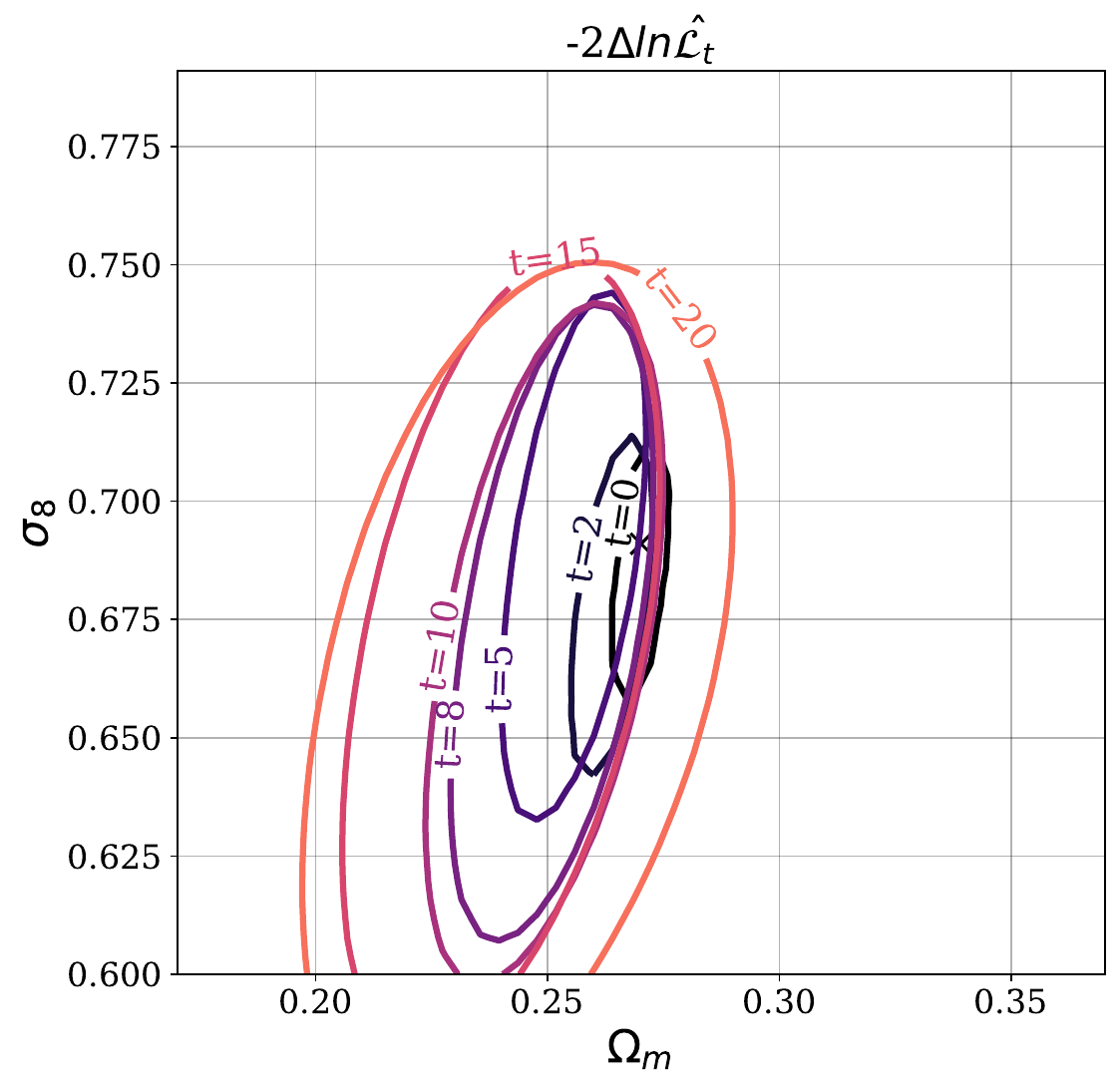}
\includegraphics[width=0.35\linewidth]{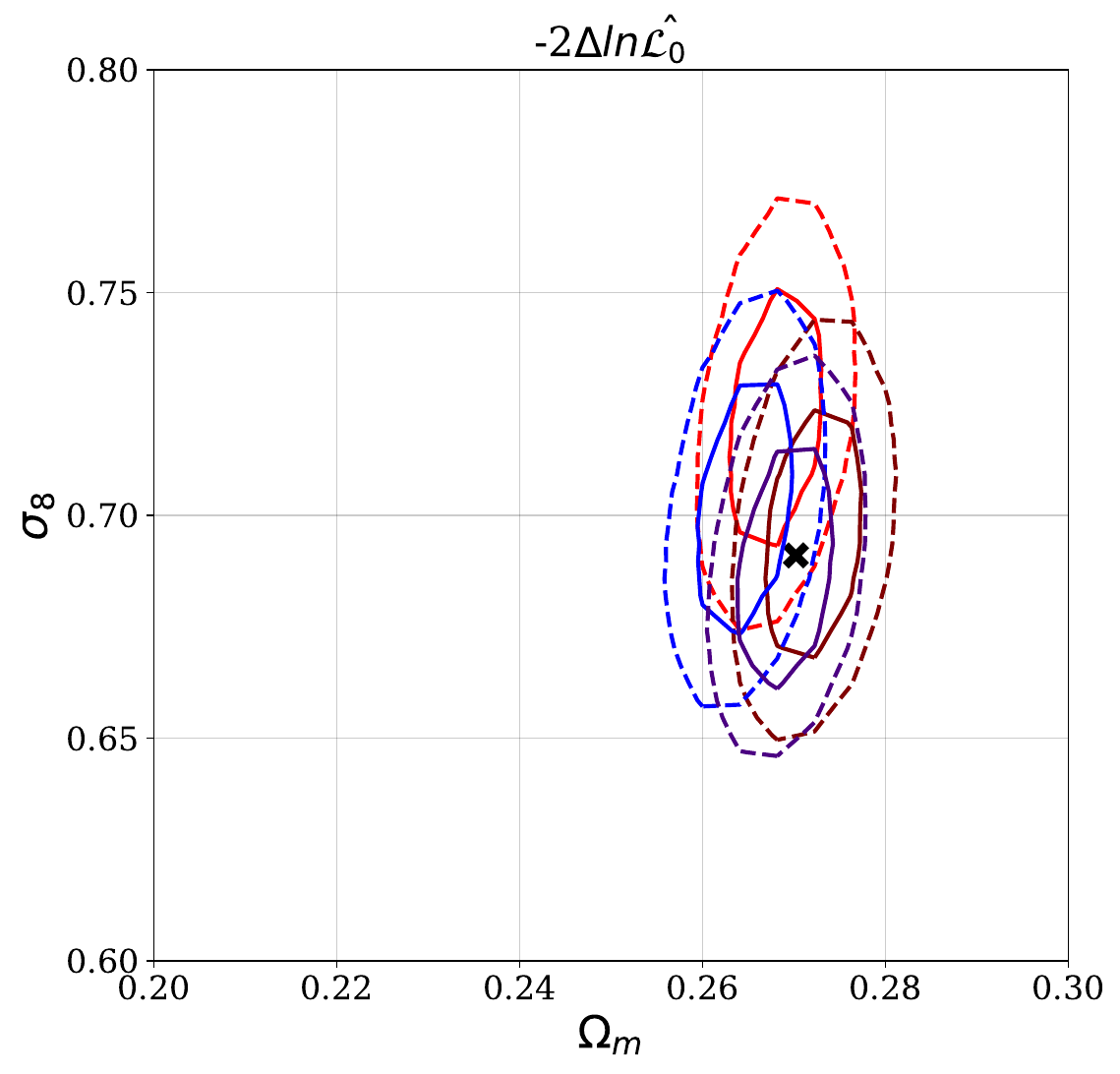}
\includegraphics[width=0.35\linewidth]{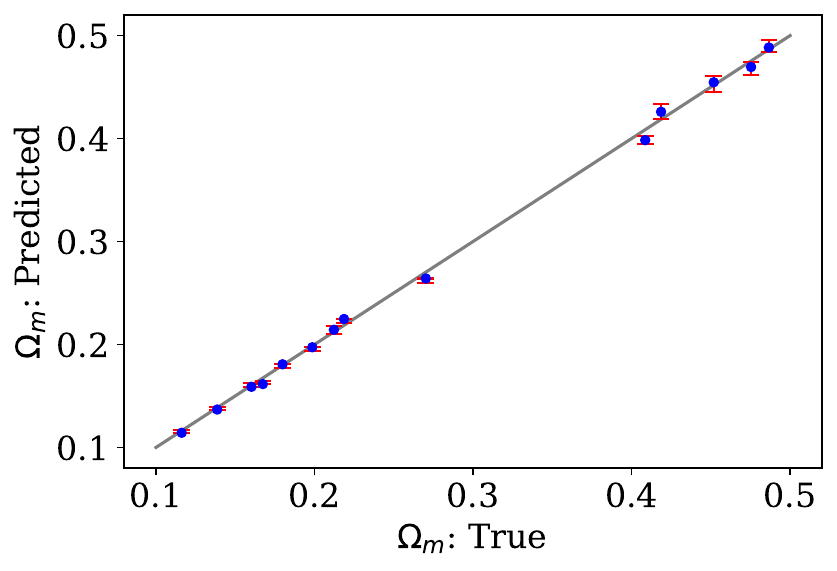}
\includegraphics[width=0.35\linewidth]{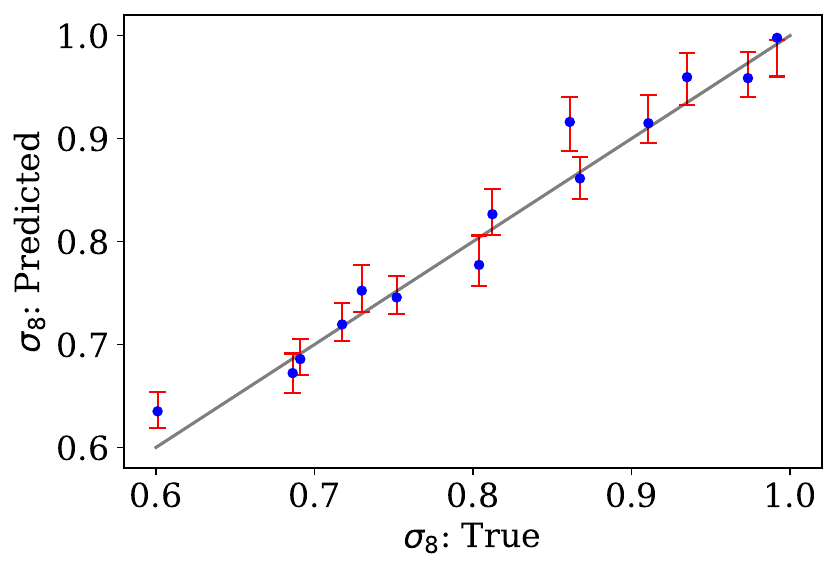}
\includegraphics[width=0.35\linewidth]{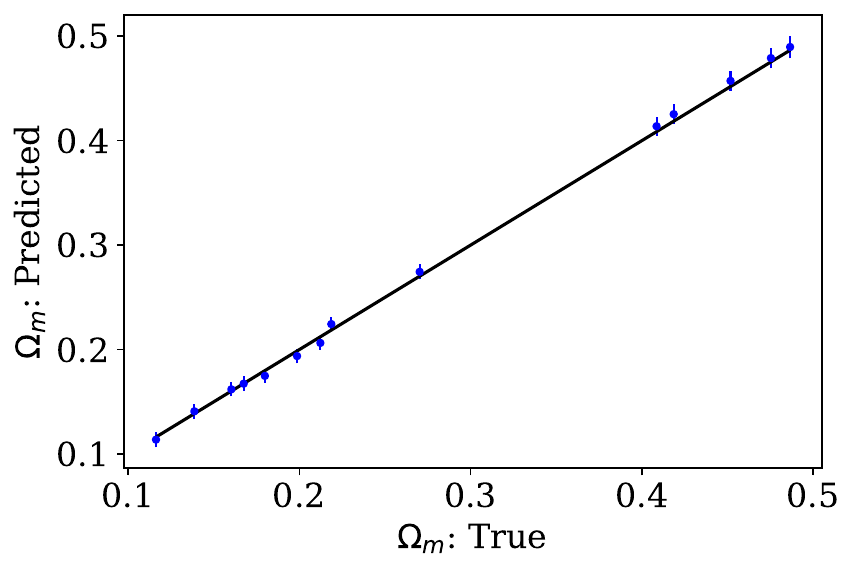}
\includegraphics[width=0.35\linewidth]{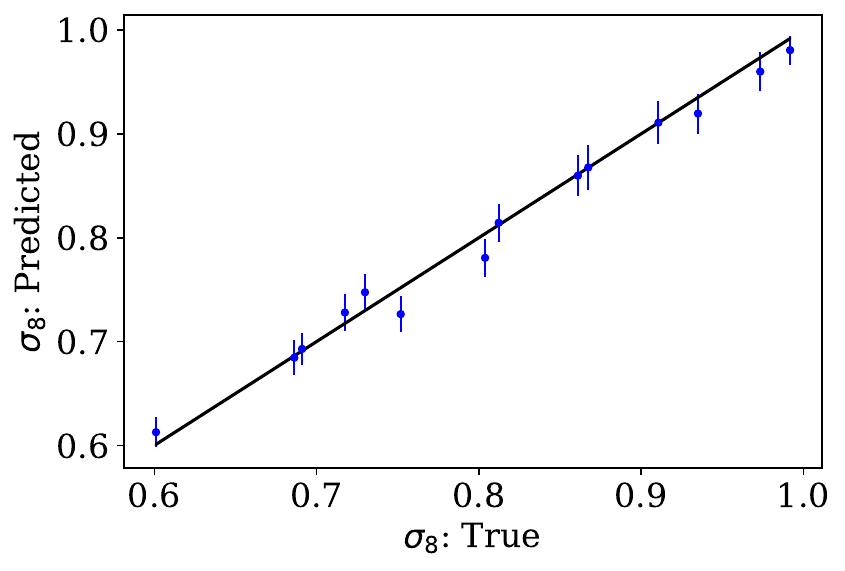}
\caption{\textbf{First Row, Left:} One sigma contours for $-2\Delta ln \hat{\mathcal{L}}$ contribution for 7 different timesteps. \textbf{First Row, Right:} One (solid) and two (dashed) sigma contours for the $-2\Delta ln \hat{\mathcal{L}}$ term for the 0th timestep, for 2 different seeds and 2 different samples of the input fields for a specific parameter. The cross demarcates the true parameter corresponding to the input fields. \textbf{Second Row:} Predicted parameter and ground truth parameter for 14 different validation parameters. The error bars correspond to the 68\% interval for the marginal probability distributions for each parameter for each input field using the $-2\Delta ln \hat{\mathcal{L}_0}$ term. \textbf{Third Row:} Predicted parameter and ground truth parameter for 14 different validation parameters using the parameter inference networks in \cite{villaescusa2021multifield}.}
\label{fig:clik_t}
\end{figure}

\section{Parameter Inference}
Diffusion models allow an evaluation of the lower bound on the log likelihood, the variational lower bound (VLB). For a conditional diffusion model, $L_{vlb} = L_0 + L_1 ... L_{T-1} + L_T = -\mathrm{log} p_\phi(x_0|\theta, x_1) + \sum_{t>1} D_{KL} [q(x_{t-1} | x_t, x_0)|| p_\phi(x_{t-1}|x_t, \theta)] + D_{KL} [q(x_T | x_0)|| p(x_T)]$  where $p_\phi$ are the learned reverse distributions parameterized by the trained neural network $\phi$ conditional on an input cosmological parameter vector $\theta$ and $q$ are the forward (analytical) distributions. \change{Since the diffusion model is conditioned on an input parameter, we can derive an upper bound on the negative log likelihood, conditional on an input parameter. One can thus investigate whether the variational lower bound terms of a trained diffusion model can be interpreted as a statistic that is sensitive to the parameter corresponding to a given input field.} For an input field $x_0$, we evaluate the $L_{t}(x_0|\vec{\theta}_{\textsc{Eval}})$ terms over a grid in $[\Omega_m, \sigma_8]$, centered on the value of the true field $\vec{\theta}_{\textsc{True}}$. The grid spans $max(\Omega_{m, \textsc{True}} - 0.1, 0.1)$ and $min(\Omega_{m, \textsc{True}} + 0.1, 0.5)$ in $\Omega_m$ and likewise for $\sigma_8$, with 50 points in both dimensions, resulting in an evaluation grid with 2500 points. To map each term's contribution to a chi-squared distribution, we first subtract the minimum value of $L_{t}(x_0|\vec{\theta}_{\textsc{Eval}})$ on the grid, and multiply it by 2 to yield $-2\Delta ln\mathcal{\hat{L}}_t$ for each t. We can then identify the 1, 2, and 3 sigma contours corresponding to this estimated chi-squared distribution with 2 degrees of freedom. We plot the contours for seven such timesteps in the \change{top left panel of Figure \ref{fig:clik_t}}. Since larger time steps correspond to progressively noisier images, the first few terms possess the most discriminatory power. \change{Since the change in $L_{vlb} (x_0|\vec{\theta}_{\textsc{Eval}})$ from one value of $\vec{\theta}_{\textsc{Eval}}$ to another is dominated by the $L_0$ term, we approximate the likelihood ratio by the contribution arising from the ratio of the first term $L_0$. We defer a more rigorous examination of the optimum subset of timesteps to use to optimize the tradeoff between faster computation and higher precision to future work.}  Next, we test whether the approximation to $-2\Delta ln \mathcal{\hat{L}}$ computed using only $L_0$ is minimized at $\vec{\theta}_{\textsc{True}}$, the true value of the parameter corresponding to the input field $x_0$. There are two sources of stochasticity that contribute to this test (see top right, Figure \ref{fig:clik_t}): the seed used to sample $x_{t=1}$ in the $L_{0}$ term, and the choice of input field sample from the true distribution for the same $\vec{\theta}_{\textsc{True}}$ (cosmic variance). In the second row, we plot the true and predicted cosmological parameters for 14 different parameters from the validation set for a single sample and seed for each parameter  using only the $-2\Delta ln \mathcal{\hat{L}}_0$ term. We convert $-2\Delta ln \mathcal{\hat{L}}_0$ to a likelihood and sum over each axis to derive a marginal likelihood for each parameter on the grid. The `predicted' parameter is the parameter at which the marginal probability distribution is maximized on the 1D grid. The one-dimensional error bars, are obtained by finding the 68\% confidence interval for each marginal. Note, that since the marginals do not account for the covariance, $\vec{\theta}_{\textsc{True}}$ might lie within the one-dimensional 1 sigma interval without lying within the two-dimensional 1 sigma contour. The constraints on $\Omega_m$ are much tighter than those on $\sigma_8$. This is \change{consistent with} and comparable to the performance of the parameter inference networks in \cite{villaescusa2021multifield} (third row). The predicted $\vec{\theta}_{\textsc{True}}$ is close to the true value of $\vec{\theta}_{\textsc{True}}$ over a broad range of parameters. Note, the 1D error bars in \cite{villaescusa2021multifield} are not derived from a two dimensional approximate likelihood, but are derived from a neural network that is trained to return the mean and the standard deviation of each parameter. Since the diffusion model's noise prediction loss is related to the reweighted VLB \citep{kingma2023understanding}, the terms of the VLB computed using the conditional diffusion model encode dependencies on the cosmological parameters. The error bars on $\Omega_m$ often do not account for the error in the prediction, and we intend to explore whether including more VLB terms yields better calibrated error bars. It would also be useful to plug the conditional diffusion model-based approximate likelihood ratios into an MCMC sampler \citep{hermans2020likelihood} and compare the constraints we obtain to those obtained from canonical summary statistics such as the power spectrum.

\section{Conclusion}
In this work, we deployed a diffusion generative model as an emulator for log cold dark matter density fields, and as a parameter inference model that can yield tight constraints on cosmological parameters. Future work could be focused on finding ways to accelerate the generation process. We also intend to examine the statistics of the fields in regular (non-log) space, convergence with respect to the number of seeds used to estimate the likelihood-based constraints and the subset of terms needed for the inference constraints to enhance the calibration of our parameter inference step. It would be interesting to compare the \change{performance} of the parameter inference step on the true fields vs the generated fields, fields from the CAMELS simulation suite with a different choice of astrophysical feedback, such as SIMBA, and the effect of adding distortions to the image. Parameter inference approaches that are able to marginalize over astrophysical feedback and observational noise are desirable.

\section{Acknowledgements}
We thank Yueying Ni, Core Francisco Park, Shuchin Aeron, Francisco Villaescusa-Navarro, Andrew K. Saydjari, and Justina R. Yang for helpful discussions. This work was supported by the National Science Foundation under Cooperative Agreement PHY2019786 (The NSF AI Institute for Artificial Intelligence and Fundamental Interactions). 


\bibliographystyle{unsrtnat}
\bibliography{sample}

\section{Appendix}
\subsection{Additional Summary Statistics}
\begin{figure}[h]
  \centering
\includegraphics[width=0.85\linewidth]{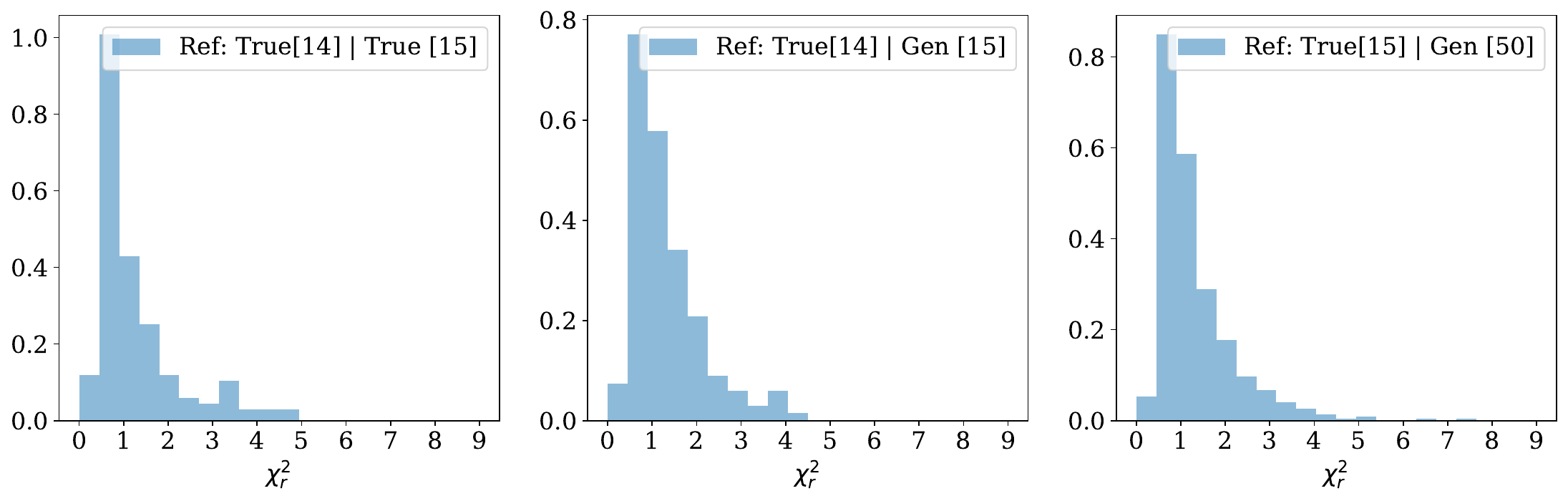}
\caption{Distribution of the reduced chi-squared statistic for the power spectra of the generated fields relative to the true fields belonging to the same parameter for 10 different validation parameters. \textit{Left, Center:} `Leave-one-out' method: 14 true fields' power spectra serve as the reference distribution and we test on the power spectra of either the remaining true field (left) or a generated field (center). We use 15 test fields per parameter for both cases and have 150 datapoints in each. \textit{Right:} All 15 true fields' power spectra serve as the reference distribution and we test on all 50 generated fields. Since there are 50 fields and 10 parameters, we have 500 datapoints in this plot.}
\label{fig:rchisq}
\end{figure}

We examine the consistency of the pixel histograms of the fields for the 3 validation parameters in Figure \ref{fig:pk}. The model appears to be slightly biased toward producing fields with higher density for the same parameter. We hope to find ways to mitigate this artefact in future work. In Figure \ref{fig:rchisq}, we plot the histograms of the distributions of the reduced chi-squared statistics computed as described in Section \ref{sec:training}. The plots on the left and the center serve as an `apples-to-apples' comparison since they use the same reference distributions and are tested on the same number of fields (the remaining true field or a generated field). While the plot with the reduced chi-squared statistic for 500 fields has some outliers, we find that the three distributions are in good agreement with each other.
\begin{figure}[h]
  \centering
\includegraphics[width=0.28\linewidth]{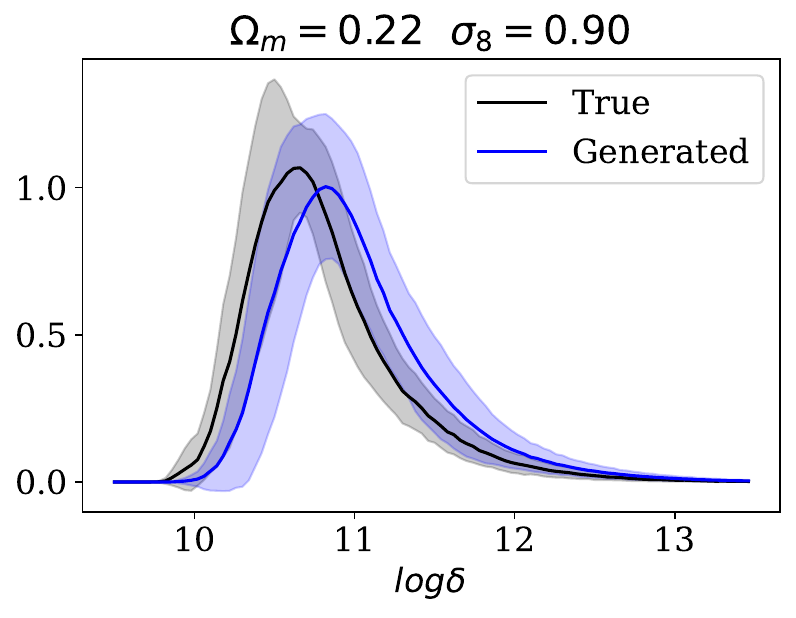}
\includegraphics[width=0.28\linewidth]{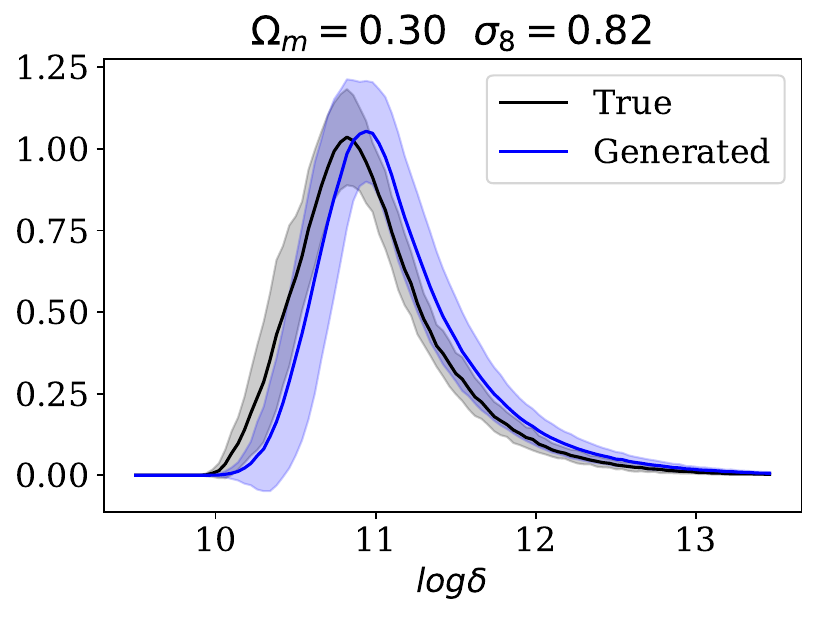}
\includegraphics[width=0.28\linewidth]{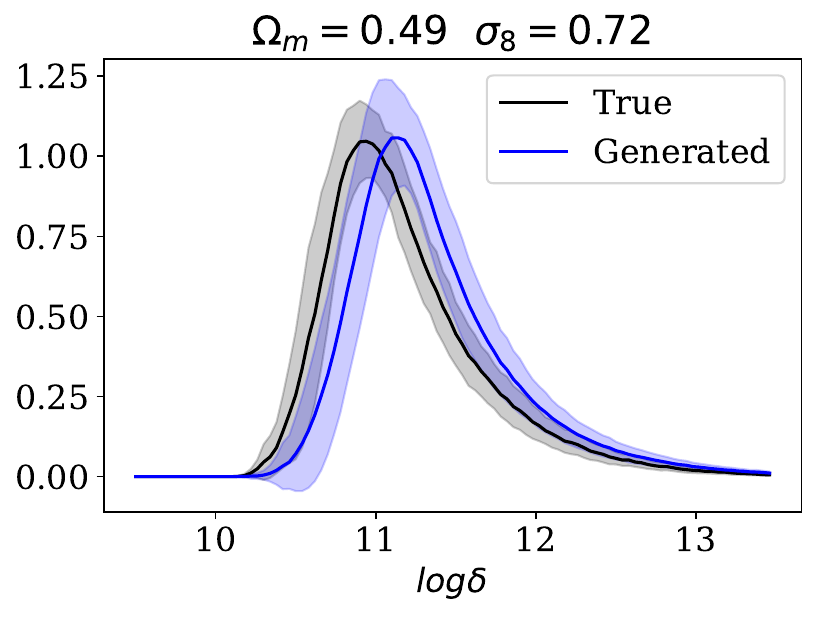}
\caption{Density histograms plotting the mean and the standard deviation envelope for 15 true fields and 50 generated fields for each parameter.}
\label{fig:hist}
\end{figure}

\subsection{Parameter Inference}
\begin{align}
& \mathbb{E}_q[-\log p_\phi(x_0|\theta_{\textsc{Eval}})] \leq \mathbb{E}_q [\rm{ D_{KL}} [q(x_T|x_0) || p(x_T)] +  \\ 
& \sum_{t>1} \rm{ D_{KL}} [q(x_{t-1}|x_t, x_0) || p_\phi (x_{t-1}|x_t, \theta_{\textsc{Eval}})] \nonumber - \log p_\phi (x_0|x_1, \theta_{\textsc{Eval}})] \\
& \textrm{For t=0}, -2\Delta ln \mathcal{\hat{L}}_0(x_0|\theta_{\textsc{Eval}}) \simeq 2[L_0 - \rm{argmin}_{\theta_{\textsc{Eval}}} L_0], \text{ where } L_0 =  - \log p_\phi (x_0|x_1, \theta_{\textsc{Eval}}) \nonumber\\
& \textrm{For t} \in [1, T-1],-2\Delta ln \mathcal{\hat{L}}_t(x_0|\theta_{\textsc{Eval}}) \simeq 2[L_t - \rm{argmin}_{\theta_{\textsc{Eval}}} L_t] \\ \nonumber
& \text{ where }  L_t = D_{KL}[q(x_{t-1}|x_t, x_0) || p_\phi (x_{t-1}|x_t, \theta_{\textsc{Eval}})] 
\end{align}
We intend to explore a more precise approximation to the VLB using the sum of a subset of the $L_t$ terms. Here, the minimum value over the grid would be computed over the sum of the terms.


\end{document}